\documentclass[aip,unsortedaddress,amsmath,preprint,endfloats*]{revtex4-1} %,linenumbers
\usepackage{graphicx}
\usepackage{bm}
\usepackage{makeidx}
\makeindex
\begin{document}
%\preprint{}

\title{Electrical and terahertz magnetospectroscopy studies of laser-patterned micro- and nanostructures on InAs-based heterostructures}

\author{O.~Chiatti}
\email[E-mail: ]{Olivio.Chiatti@physik.hu-berlin.de}
%\homepage[Website:]{http://www.physik.hu-berlin.de/gnm}
%\thanks{}

\author{S.~S.~Buchholz}
\affiliation{Novel Materials Group, Institut f\"{u}r Physik, Humboldt-Universit\"{a}t zu Berlin, 12489 Berlin, Germany} %AG Neue Materialien,

\author{Ch.~Heyn}
\author{W.~Hansen}
\affiliation{Institut f\"{u}r Angewandte Physik, Universit\"{a}t Hamburg, 20148 Hamburg, Germany} %FG Wachstum,

\author{M.~Pakmehr}
\author{B.~D.~McCombe}
\affiliation{Department of Physics, University at Buffalo, the State University of New York, Buffalo, NY 14260, USA}

\author{S.~F.~Fischer}
\affiliation{Novel Materials Group, Institut f\"{u}r Physik, Humboldt-Universit\"{a}t zu Berlin, 12489 Berlin, Germany} %AG Neue Materialien,


\date{\today}

\begin{abstract}
Nanostructures fabricated from narrow-gap semiconductors with strong spin-orbit interaction (SOI), such as InAs, can be used to filter momentum modes of electrons and offer the possibility to create and detect spin-polarized currents entirely by electric fields.
Here, we present magnetotransport and THz magnetospectroscopy investigations of Hall-bars with back-gates made from in InGaAs/InAlAs quantum well structures with a strained 4~nm InAs-inserted channel.
The two-dimensional electron gas is at 53~nm depth and has a carrier density of about $6\times10^{11}$~cm$^{-2}$ and mobility of about $2\times10^{5}$~cm$^2$/Vs, after illumination. Electrical and THz optical transport measurements at low temperatures and in high magnetic fields reveal an effective mass of 0.038$m_{0}$ and an anisotropic $g$-factor of up to 20, larger than for bulk InAs or InAs-based heterostructures.
We demonstrate that quasi-one-dimensional channels can be formed by micro-laser lithography.
The population of subbands is controlled by in-plane gates.
Contrary to previous reports symmetric and asymmetric in-plane gate voltages applied to quasi-one dimensional channels did not show indications of SOI-induced anomalies in the conductance.
\end{abstract}


\maketitle

%%\section{\label{sec:intro}Introduction}
Conventional electronics manipulates charges using electric fields and macroscopic magnetization of ferromagnetic materials.
The field of spintronics promises novel devices that go beyond the conventional paradigm, by manipulating at the same time spin and charge of electrons.~\cite{zutic-2004-rmp}
One of the biggest challenges in this field is the creation, detection and manipulation of spin-polarized currents.
A possible tool for an all-electrical control of spin-polarized currents is the spin-orbit interaction (SOI).
In materials lacking inversion symmetry, the SOI is one of the mechanisms coupling the momentum of the electrons with their spin, creating correlations between spin polarization and charge current.~\cite{silsbee-2004-jpcm,wu-2010-prep}

Inversion asymmetry is typically either the result of the bulk crystal structure, e.~g., in zinc-blende III-V or II-VI semiconductors, or of an asymmetric structure, e.~g. in a junction of different materials, %of the material itself or a combination of materials,
such as semiconductor heterostructures.
The former case is commonly referred to as bulk inversion asymmetry (BIA) or the Dresselhaus effect, and the latter as structural inversion asymmetry (SIA) or the Rashba effect.~\cite{silsbee-2004-jpcm,wu-2010-prep,zhang-2014-np}
Each leads to an anisotropic SOI.
III-V semiconductor heterostructures and quantum wells (QWs) can have SOI effects contributed from both mechanisms.
Further confinement in 1D structures, especially with lateral gates, can be used to define a dominant direction of the anisotropic SOI.~\cite{zutic-2004-rmp,silsbee-2004-jpcm,wu-2010-prep}
A pair of lateral gates combined with a quantum point contact (QPC) can also create a strong SIA in the plane of the 2D electron gas (2DEG), and an effective Rashba magnetic field normal to the plane, whose strength is controlled by the gate voltage difference and whose direction can be flipped by changing the sign of the lateral potential difference.~\cite{debray-2009-nn}

Narrow-gap semiconductors with large SOI, such as InAs, have received attention as potential building blocks for spintronic devices,~\cite{silsbee-2004-jpcm,wu-2010-prep} in particular as QW structures.~\cite{richter-2000-apl,grundler-2000-prl}
Semiconductor heterostructures offer the opportunity of using well-developed techniques for the fabrication of low-dimensional electron systems, with great flexibility in manipulating charges and the momentum of electrons by electric fields and nanopatterning.~\cite{davies-1998,beenakker-1991-ssp}
Unresolved issues exist on the magnitude of spin-splitting in magnetic fields and its correlation to confinement.
And, whether the electric field perpendicular to the electron momentum~\cite{debray-2009-nn} or the structural asymmetry\cite{richter-2000-apl} plays the main role in the reported SOI-induced effects and if it can be used to invoke and control electrically spin-polarized transport in low dimensions.

Here, we investigate the $g$-factor and SOI-induced phenomena in Hall-bars with back-gates and QPC structures with in-plane and back-gates, fabricated with micro-laser photolithography in shallow inverted InGaAs/InAlAs QWs with an inserted InAs layer.
The low-temperature magnetotransport of the Hall-bars shows Shubnikov-de Haas (SdH) oscillations and quantum Hall effect (QHE), and is used to determine carrier density $n$ and mobility $\mu$.
THz photoresponse and magnetotransmission measurements have been used to determine the effective mass $m^*$ and the Land\'e factor $g$.
The QPCs show the effects of conductance quantization at low temperatures and the effect of transverse electrical fields is investigated by asymmetric in-plane gate voltages.

%%\section{\label{sec:expdetails}Experimental details}
The wafer was grown by molecular beam epitaxy and contains a In$_{0.75}$Ga$_{0.25}$As/In$_{0.75}$Al$_{0.25}$As QW, with a strained, 4~nm thick InAs layer inserted about 53~nm below the surface.~\cite{heyn-2003-jcg,stroppa-2005-prb}
The 2D electron gas (2DEG) is localized in the narrow InAs QW.
Initial characterization of the wafer at 4.2~K in the dark yields electron densities in the range $n \approx (2-3){\times}10^{11}$~cm$^{-2}$ and mobilities in the range $\mu \approx (2-9){\times}10^{4}$~cm$^{2}$/Vs.
Under continuous light exposure conditions, %(in the optical magnet cryostat),
the density increases  by roughly a factor of two and the mobility from the SdH scattering time increases up to about $3\times10^{5}$~cm$^{2}$/Vs.

The Hall-bars were fabricated by micro-laser photolithography and wet-chemical etching to form the mesa.
Atomic force microscopy shows that the etching process removes material down to a depth of approximately 90~nm, below the doping layer.~\cite{heyn-2003-jcg}
The contacts to the 2DEG were made by sputter-deposition of a 5~nm layer of chromium or titanium, followed by a 50~nm layer of gold, without annealing.

\begin{figure}[tb]
    \includegraphics[width=0.85\columnwidth]{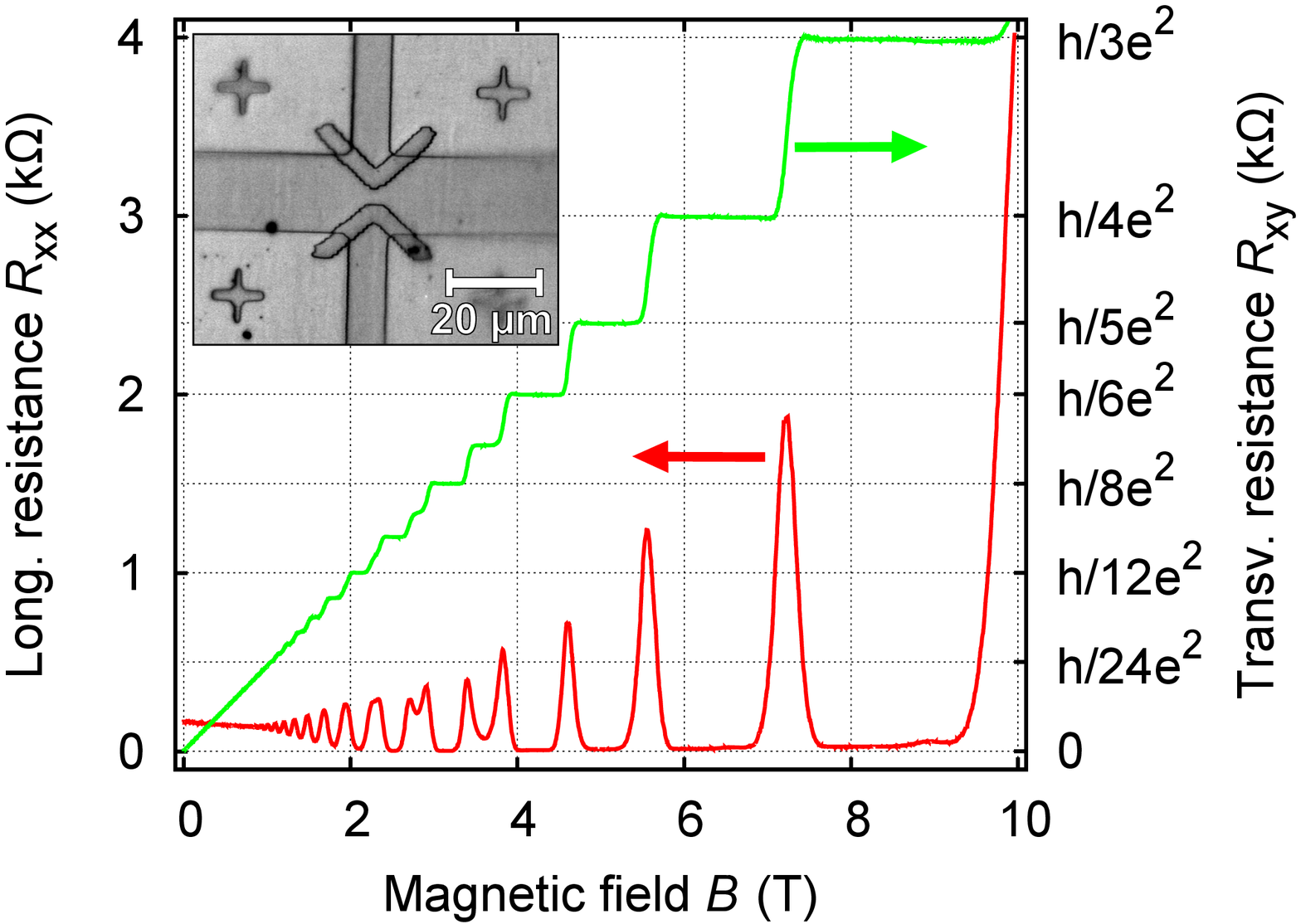}
    \caption[Magnetotransport]{\label{fig:hall-bars_mt}%
        Low-temperature magnetotransport measurements of Hall-bar HB1, after illumination with an infrared light-emitting diode.
        The longitudinal resistance $R_{xx}$ and the transversal resistance $R_{xy}$ were measured simultaneously as a function of perpendicular magnetic field $B$.
        The measurement was performed at a bath temperature $T_{bath} = 0.3$~K, using a lock-in technique with current $I = 20$~nA.\\
        The inset shows a confocal microscopy image of sample QPC1.
        The dark grey areas are the mesa, the two V-shaped marks are the etched trenches that form the constriction of the QPC.
        The lithographic width and length of the gap between the trenches are approximately 3.5~${\mu}$m and 3~${\mu}$m, respectively.
    }
\end{figure}

The QPCs were fabricated with high-resolution micro-laser photolithography and wet-chemical etching as pairs of trenches, shaped like facing 'V's and about 3~$\mu$m wide, separating the 2DEG of a Hall-bar in three electrically isolated regions, as shown in Fig.~\ref{fig:hall-bars_mt}.
The samples were mounted on chip-carriers and contacted by wedge-bonding with gold or aluminum wire.
The bottom of the chip-carrier is used as a back-gate.

Electric transport measurements were performed with a $^3$He-cryostat in a 10~T magnet.
The sample was mounted in vacuum at the end of a copper 'cold-finger', with the 2DEG perpendicular to the magnetic field.
All the measurements used lock-in techniques for signal recovery.
Optical transport measurements were performed in a 10~T optical access magnet system, equipped with a variable-temperature insert.
An optically pumped THz laser was used as the radiation source.
The sample was mounted at the center of the magnetic field, on a holder that allowed full 360$^{\circ}$ rotation with an accuracy of about 0.5$^{\circ}$, for the tilted field measurements.
The sample holder was immersed in superfluid helium at a bath temperature of $T_{bath} = 1.4 - 1.6$~K.

%%\section{\label{sec:results}Experimental results}
%%\subsection{\label{ssec:fabric}Fabricated Hall-bars and QPC structures}

\begin{figure}[tb]
    \includegraphics[width=0.99\columnwidth]{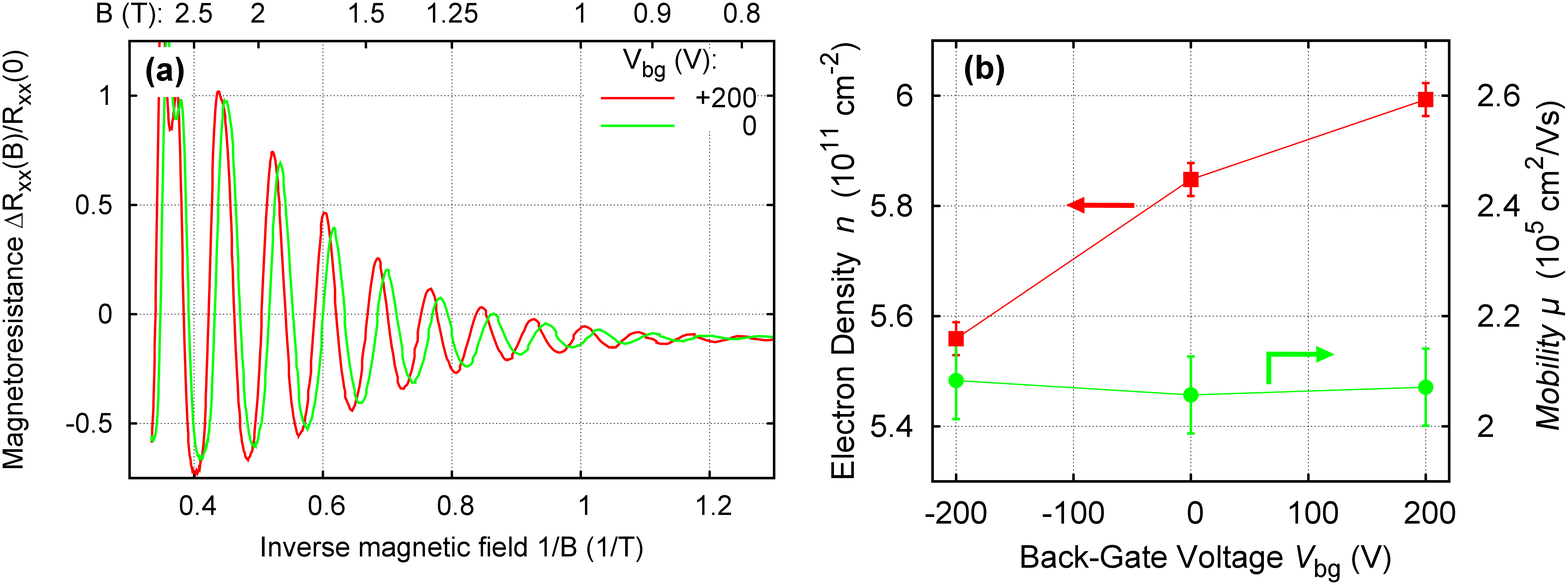}
    \caption[Back-gate voltage]{\label{fig:hall-bars_sdh}%
        Effect of back-gate voltage in Hall-bar HB1:
%        \begin{enumerate}[(a)]
%            \item
        (a) SdH oscillations in the resistance, ${\Delta}R_{xx}/R_{xx}(0) = \left(R_{xx}(B)-R_{xx}(B=0)\right)/R_{xx}(B=0)$, for two values of back-gate voltage $V_{bg}$: 0~V (green) and +200~V (red).
        Temperature $T_{bath}$ and current $I$ are the same as in Fig.~\ref{fig:hall-bars_mt}.
%            \item
        (b) Electron density $n$ and mobility $\mu$, as obtained from SdH oscillations.
        The values are in agreement with the carrier density obtained from the low-field Hall resistance.
        The straight lines between data points are guides to the eye.
%        \end{enumerate}%
    }
\end{figure}

The inset of Fig.~\ref{fig:hall-bars_mt} shows a confocal microscopy image of a QPC (sample QPC1), etched into a Hall-bar.
The lithographic width and length of the gap between the trenches are approximately 3.5~${\mu}$m and 3~${\mu}$m, respectively.
Using carrier density and mobility values obtained from the magnetotransport measurements of Hall-bars (Fig.~\ref{fig:hall-bars_sdh}), %(sec.~\ref{ssec:hall-bars:n-mu}),
we estimate the Fermi wavelength to be  $\lambda_F \approx 33$~nm and the mean free path for $T \lesssim 1.5$~K to be $L_e \gtrsim 2.6$~${\mu}$m.%

%%\subsection{\label{ssec:hall-bars:mt}Magnetotransport of  Hall-bars}
Figure~\ref{fig:hall-bars_mt} shows low-temperature magnetotransport measurements of a Hall-bar (sample HB1), after illumination with an infrared light-emitting diode.
The measurements were performed at a bath temperature of $T_{bath} = 0.3$~K, with the magnetic field $B$ perpendicular to the 2DEG.
The longitudinal resistance $R_{xx}$ and the transversal resistance $R_{xy}$  were measured simultaneously by lock-in technique, with a current $I = 20$~nA.

$R_{xx}$ shows SdH oscillations between $B \approx 0.7$~T and $B \approx 2$~T, and at higher fields has flat minima.
$R_{xy}$ shows plateaus at the same fields where $R_{xx}$ has minima.
This behavior is consistent with the formation of well-defined Landau levels (LLs) and the observation of the QHE for $B > 2$~T.
Before illumination the minima in $R_{xx}$ are not zero and the plateaux in $R_{xy}$ are not at their theoretical values, $R_{xy} = \frac{1}{n}\frac{h}{e^{2}}$.
This indicates a conduction channel parallel to the 2DEG.~\cite{heyn-2003-jcg}
After illumination, the parallel conduction vanishes and the electron density in the 2DEG is increased.

Beating patterns in SdH oscillations are indicators of the presence of SOI-created effective magnetic fields.~\cite{park-2011-tsf}
However, they become less pronounced with decreasing carrier density.~\cite{nitta-1997-prl}
In our measurements beating patterns are not observed, indicating that the samples are in the low-density range.~\cite{moller-2003-apl}
Also, the mobility may not be high enough to permit the observation of many SdH oscillations at the requisite low magnetic fields.
However, estimates of the broadening of the Landau levels around $B = 0.7$~T (where SdH oscillations are resolved) yield values around 0.5~meV (for $\tau_{SdH} = 0.1 \tau_e$) and around $B = 2.5$~T (where spin-splitting is resolved) the LL broadening is around 0.6~meV (for $g = 20$).
That gives us an upper limit for the Rashba parameter of $\alpha_R < 10^{-10}$~eVcm.

%%\subsection{\label{ssec:hall-bars:opt}Terahertz magnetospectroscopy of Hall-bars}
\begin{figure}[tb]
    \includegraphics[width=0.95\columnwidth]{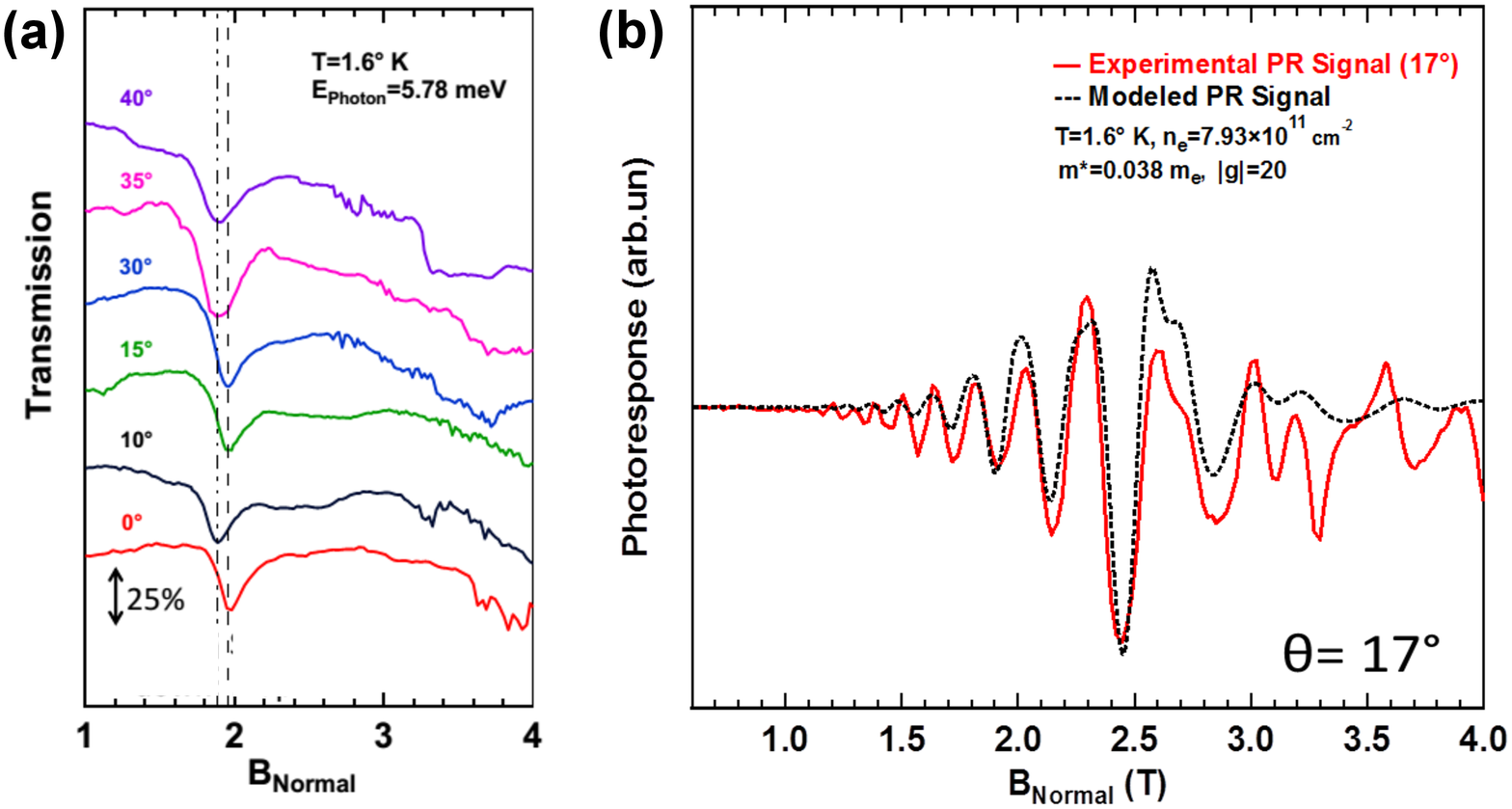}
    \caption[Optical transport]{\label{fig:hall-bars_opt-mt}%
        THz magnetospectroscopy of Hall-bar HB2:
%        \begin{enumerate}[(a)]
%            \item
        (a) Magnetotransmission as a function of the magnetic-field component normal to the 2DEG, $B_{Normal} = B \cos(\theta)$, at varying angles $\theta$ between external magnetic field $B$ and the 2DEG.
        There is a sharp line at $B_{Normal} \approx 1.9$~T corresponding to cyclotron resonance (CR) absorption.
%            \item
        (b) Photoresponse (PR) signal at angle $\theta = 17^{\circ}$: it is approximately the difference between $R_{xx}(B,T(B))$ with laser on and $R_{xx}(B, T_{bath})$ with laser off, where $T(B)$ with laser on is a temperature profile dependent on the CR absorption line.
        The signal is modelled with a theoretical curve  for 2D SdH oscillations that takes into account CR heating of the electrons, which modulates the amplitude of the SdH oscillations.
        The fit yields the electron density $n$ from the period of oscillations, the effective mass $m^*$ from the amplitude envelope, the $g$-factor from the splitting of the LLs, and the SdH scattering time $\tau_{SdH}$ from the overall onset of the oscillations and their general growth with $B$.
%        \end{enumerate}%
    }
\end{figure}

High frequency (THz) magnetospectroscopy of the conductivity of quasi-2D and -1D structures provides an excellent characterization tool to probe important parameters of these electron gases, such as the effective mass $m^*$ and the $g$-factor, and complement dc magneto-transport studies.
Also, angle-dependent measurements of photocurrents induced by far-infrared radiation have been used to determine the relative strengths of the Rashba and Dresselhaus contributions in InAs-based QWs.~\cite{ganichev-2004-prl}
Figure~\ref{fig:hall-bars_opt-mt} shows the results of our THz optical studies on a Hall-bar (sample HB2).
The magnetotransmission (Fig.~\ref{fig:hall-bars_opt-mt}a) was performed on an unpatterned region and shows a sharp minimum at $B = 1.9$~T, corresponding to electron cyclotron resonance (CR).
The origin of the broader transmission minimum at higher fields is presently uncertain, although it may result from CR absorption in the barrier doping region, where carrier have a larger effective mass, or from neutral donor magneto-optical transitions ($1s-2p$, $M = -1$).
The magnetic field position of the sharp line and the laser photon energy of 5.83~meV yields the cyclotron effective mass of the carriers: $m^* = (0.0386 \pm 0.0005) m_0$, a value consistent with previous measurements in similar structures.~\cite{richter-2000-apl}

The photoresponse (PR) signal of Hall-bar HB2 is shown in Fig.~\ref{fig:hall-bars_opt-mt}b along with a fit discussed below.
The PR in this field/photon energy region results from CR heating of electrons.
Thus the data are approximately the difference between $R_{xx}(B,T(B))$ with laser on and $R_{xx}(B, T_{bath})$ with laser off, where $T(B)$ with laser on is taken to be a Lorentzian temperature profile directly reflective of the CR absorption line, and $T_{bath}$ is the constant bath temperature with laser off.
The PR signal shows the resonant absorption primarily as an envelope of amplitude modulation of the SdH oscillations.

We have modelled the PR signal by using the 2D SdH expression~\cite{isihara-1986-jpc} and summing over several SdH harmonics.
Fitting the PR yields the electron density $n$ from the period of oscillations, the effective mass $m^*$ from the amplitude envelope, the $g$-factor from the splitting of the LLs, and the SdH scattering time $\tau_{SdH}$ from the overall onset of the oscillations and their general growth with $B$.
The data of Fig.~\ref{fig:hall-bars_opt-mt} is complex because of the additional absorption near CR at higher fields, and we found it necessary to include in the model this absorption which also heats the electrons.
Fitting the data yields $g \approx 20$ and $m^* \approx 0.038 m_0$, close to that obtained from the magnetotransmission, which is considerably more accurate.
The values of $n$ and $\mu$ are close to those extracted from dc magnetotransport measurements.

It is common to use the so-called ``coincidence method'' in concert with the cyclotron effective mass to determine the $g$-factor from magnetotransport measurements.~\cite{fang-1968-pr} %
We have used this method and find that the angle at which the SdH period doubling occurs is $\theta = (79.0 \pm 1.5)^{\circ}$, from which the $g$-factor determined in this manner is only $|g| = (5.0 \pm 0.8)$, with the measured effective mass of $m^* = (0.0385 \pm 0.0006) m_0$.
A similar value was obtained from similar measurements on nearly identical samples.~\cite{moller-2003-apl}
We attribute this large variation of the band $g$-factor with angle (factor of about three) to anisotropy created by the QW confinement;~\cite{kiselev-1998-prb}
the observed anisotropy is consistent with calculations for a similar sample~\cite{khaetskii-2014-private} incorporating wavefunction penetration into the surrounding material,~\cite{moller-2003-apl} but the magnitudes are considerably larger than these single-particle band calculations predict.
In fact the magnitude of $g$-factor for $B$ perpendicular to the plane of the QW is larger than the magnitude at the bottom of the conduction band for bulk InAs of 14.8.
We believe this enhancement is related to many-body exchange effects, and theoretical estimates~\cite{khaetskii-2014-private} yield large exchange contributions to the $g$-factor (${\Delta}g_{exch} \approx 6$).

Other investigators have also reported very large $g$-values for $B$ perpendicular to the plane of the 2DEG in 15 nm InAs/AlSb QWs.~\cite{sadofyev-2002-apl}
These authors found values that oscillated between between 15 and 60 with magnetic field, from a simple model by assuming that the $R_{xx}$ peaks occur when the LL centers are coincident with the fixed Fermi energy (this ignores self consistency of the confining potential) and with the bottom of the band effective mass of InAs ($m^* = 0.023 m_0$).  The net result is an overestimate of the $g$-factor by roughly a factor of 1.5.
The oscillations with field are attributed to a modulation of the exchange interaction by changes in screening when the Fermi energy moves from extended states to localized states.
Other early work, both experimental and theoretical,~\cite{raymond-1985-ssc,ando-1974-jpsj} also found large $g$-factor enhancement that oscillates with field in GaAs and Si 2DEGs.
Very recently, theoretical and experimental work~\cite{krishtopenko-2011-jpcm,krishtopenko-2012-sc} in InAs/AlSb heterostructures highlights the role of conduction band nonparabolicity and SOI, as well as exchange effects.
In particular, at low magnetic fields the Rashba spin-splitting increases the amplitude of $g$-factor oscillations in the presence of structural asymmetry.~\cite{krishtopenko-2011-jpcm,krishtopenko-2012-sc}
It is clear that the results and the relevant physics associated with $g$-factor measurements can be very sample dependent and method dependent and great care must be taken in quoting values.

%%\subsection{\label{ssec:hall-bars:n-mu}Determination of carrier density and mobility}

Figure~\ref{fig:hall-bars_sdh}a shows Shubnikov-de Haas (SdH) measurements  for two different back-gate voltages $V_{bg}$.
The measurements were performed on sample HB1 and in the same conditions as in Fig.~\ref{fig:hall-bars_mt}.
The back-gate voltage was applied using a \emph{Keithley} SMU~2401 as a voltage source.
The ($1/B$)-periodicity of SdH oscillations changes with varying $V_{bg}$ in a manner consistent with a change of the carrier density.
Figure~\ref{fig:hall-bars_sdh}b shows the carrier density $n$ and mobility $\mu$, respectively, extracted from these measurements.
$n$ varies between $5.55$ and $6.00 \times 10^{11}$~cm$^{-2}$, while $\mu$ is approximately constant at $2.1 \times 10^{4}$~cm$^{2}$/Vs.

Previous work on InGaAs/InAlAs QW structures with InAs-inserted channels concluded that the first two-dimensional (2D) subband is centered on the InAs channel, but the wavefunction penetrates the surrounding material.~\cite{richter-2000-apl}
Using a combination of top- and back-gate, the electrical field perpendicular to the QW could be varied at constant carrier density.
The resulting change in the SOI strength was interpreted as the result of a shift of the wavefunction between QW and barriers.~\cite{grundler-2000-prl}
Our results reported above are fully in agreement with these reports.

\begin{figure}[tb]
    \includegraphics[width=0.85\columnwidth]{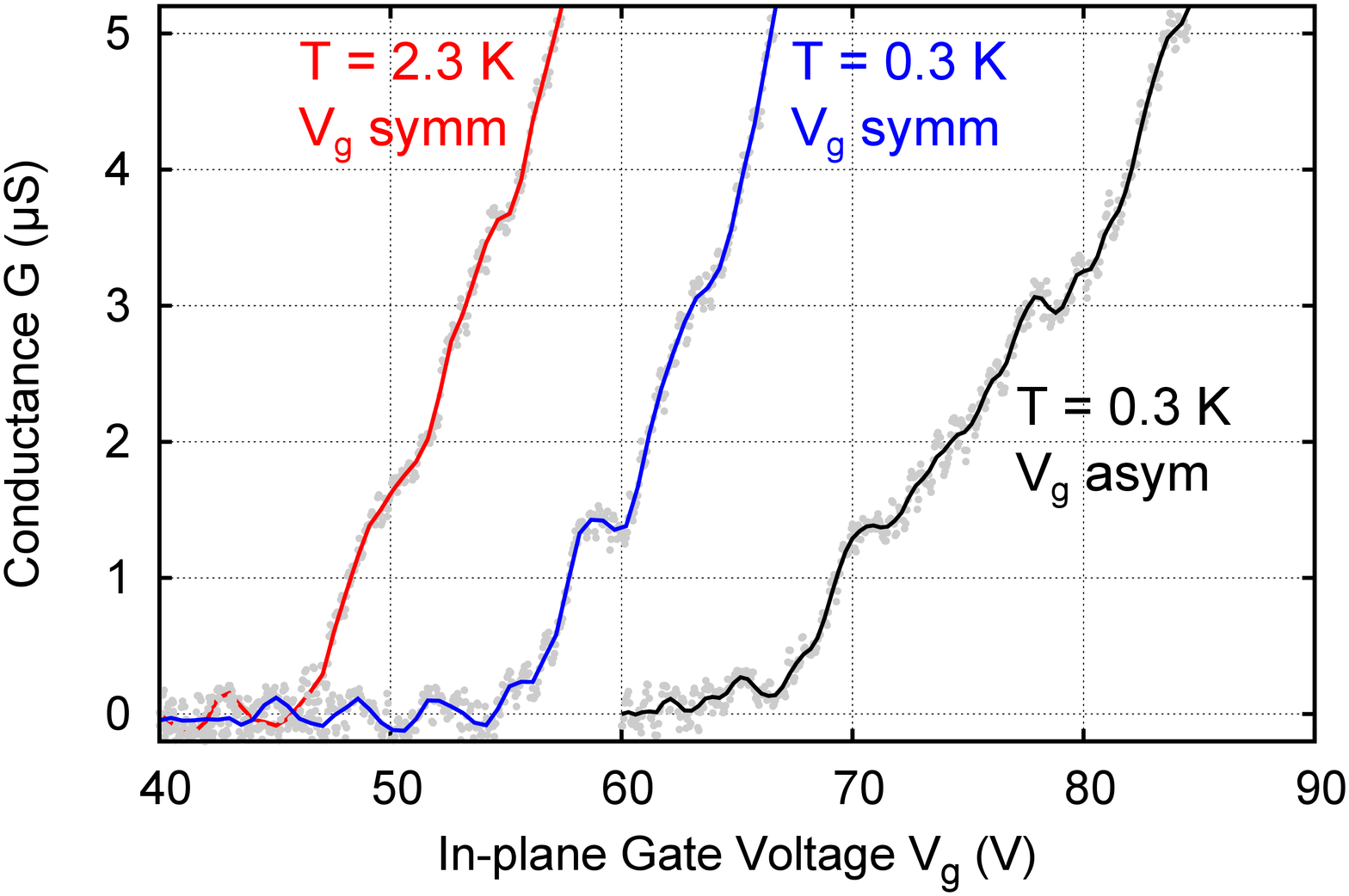}
    \caption[QPCs]{\label{fig:qpc_cond}%
        Conductance quantization:
        Two-terminal conductance $G$ in zero magnetic field of sample QPC1 as a function of in-plane gate voltage $V_g$, at two bath temperatures, $T_{bath} = 0.3$~K (blue and black) and 2.3~K (red), and with $V_{ac} = 10~{\mu}$V.
        \emph{$V_g$ symm} indicates that $V_g$ was applied symmetrically, $V_g^{left} = V_g^{right} = V_g$ (red and blue); \emph{$V_g$ asym} indicates asymmetric voltages, $V_g^{left} = V_g; V_g^{right} = 0$ (black).
    }
\end{figure}

%% QPC
%%\subsection{\label{ssec:qpcs:cond}Transport in QPCs}
In the following, our results on the laterally confined constrictions are presented.
Figure~\ref{fig:qpc_cond} shows the two-terminal conductance $G$ as a function of in-plane gate voltage $V_g$ of sample QPC1 in zero magnetic field.
$G(V_g)$  is depicted for two bath temperatures, $T_{bath} = 0.3$~K and 2.3~K, with $V_g$ applied symmetrically: $V_g^{left} = V_g^{right} = V_g$.
Both conductance curves show plateau-like structures at similar conductance values.
These values are $G_1 \approx 1.5-1.6~{\mu}$S and $G_2 \approx 3.1-3.7~{\mu}$S, respectively, and are much lower than expected for the first two plateaus in QPCs.
Usually this indicates a large series resistance $R_s$ and it is common to correct the conductance curve $G(V_g)$ for $R_s$.
In this case, if we identify $G_1$ and $G_2$ with the first two plateaus, $2e^2/h$ and $4e^2/h$ respectively, we obtain $R_{s1} \approx 650~$k${\Omega}$ and $R_{s2} \approx 300~$k${\Omega}$.
This implies that the series resistance is not constant, but depends strongly on the gate-voltage.
Four-terminal measurements, performed to avoid the resistance of the ohmic contacts, indicate that the origin of this series resistance lies in the bulk of the 2DEG.
The conductance becomes zero after pinch-off, indicating that there is no parallel conduction.

We took great care to investigate the influence of asymmetrically applied lateral electric fields, because a previous study~\cite{debray-2009-nn} reported that the first conductance plateau in an InAs-based QPC was shifted from $2e^2/h$ to $e^2/h$, when an asymmetric gate-voltage is applied to the in-plane gates.
Based on model calculations the anomaly had been attributed to the creation of a SOI-induced fully spin-polarized channel in the quasi-1D region of the QPC.~\cite{debray-2009-nn}
While in our case, the 2DEG (asymmetric QW) has about half the density and four times the mobility as reported in Ref.~\onlinecite{debray-2009-nn} the experiments compare in the lateral confinement geometry and comparable electric field magnitude ($10^4 - 10^5$~V/cm).
Our results are depicted in Fig.~\ref{fig:qpc_cond} showing  $G(V_g)$ at $T_{bath} = 0.3$~K for $V_g$ is applied symmetrical and asymmetrical ( $V_g^{left} = V_g; V_g^{right} = 0$).
It is clearly visible, that the conductance characteristics for both, symmetric and asymmetric in-plane gate-voltages, show the same quantization feature.
There is no indication of a SOI-induced conductance anomaly.
Therefore we cannot confirm that the electric field perpendicular to the momentum of electrons in a narrow or quasi-1D channel leads to observable SOI-induced effects in transport measurements.

%% DISCUSSION OF QPC
A study of  an X-shaped QPC fabricated from the same QW heterostructure as in our work showed quantized conductance steps with an overlap of universal conductance fluctuations.~\cite{lehmann-2014-sst}
A feature in the conductance at $G \approx 0.5{\cdot}G_0$, where $G_0 = \frac{2e^2}{h}$, in zero and finite magnetic fields was identified as signature of SOI-induced spin splitting.
However, our results show that conductance quantization effects are achieved even by constrictions formed by high-precision optical lithography.
Relating to the four narrow 2DEG reservoir leads attached in Ref.~\onlinecite{lehmann-2014-sst} to the QPC this might infer that the actual confinement potential therein is more complicated than that of a single QPC constriction.

%%\section{\label{sec:concl}Conclusions}
In conclusion, carrier density and mobility were determined by back-gate dependent magnetotransport from 2DEG hosted in an asymmetric InAs-based heterostructure.
Due to the low-density range the SdH oscillations in the magnetoresistance did not reveal beating patterns originating from SOI.
However estimates of the broadening of the Landau levels yield an upper limit for the Rashba parameter of $\alpha_R < 10^{-10}$ eVcm.
From THz magnetospectroscopy in tilted fields at the cyclotron resonance the effective mass of the carriers was determined.
The resulting value is in agreement with previous results.
The photoresponse shows SdH oscillations modulated in amplitude by resonant carrier heating.
A large $g$-factor of 20 was determined, while the coincidence method yields a value of 6.4.
This $g$-factor anisotropy  is interpreted as the consequence of QW confinement enhanced by many-body exchange effects.
We demonstrated that quantum point contacts can be fabricated in InAs-HEMTs, using high-resolution laser photolithography and wet-chemical etching.
In-plane gates allow to control carrier density and width of the QPC constriction.
Conductance quantization persists in a large temperature range from 0.3~K to 2.3~K.
While application of symmetric and asymmetric in-plane gate-voltages did not indicate any SOI-induced conductance anomaly as reported previously, future experiments with combined top- and in-plane gates may clarify this issue by allowing a variable electric field orientation to be applied, i.e. perpendicular to the momentum of electron in a quasi-1D channel and rotated from perpendicular to parallel to the QW.

% If in two-column mode, this environment will change to single-column format so that long equations can be displayed.
%\begin{widetext}
%$$\mbox{put long equation here}$$
%\end{widetext}

% \begin{table}
% \end{tabular}
% \end{table}

\begin{acknowledgments}
OC and SFF thank Ch. Riha for technical support.
The authors acknowledge financial support from the joint DFG and NSF ``Materials World Network'' PAK 556 grant Nr. Fi932/4-1 at the Humboldt-Universit{\"a}t zu Berlin and \#1008138 at the University at Buffalo, respectively.
OC and SFF further acknowledge partial sinancial support within the DFG Priority Program SPP~1666 grant Nr. Fi932/5-1.
CH and WH acknowledge support from BMBF project 01DJ12099 ``Era.~Net.~Rus: SpinBar''.
\end{acknowledgments}

%\bibliography{OChiatti_InAs-HEMT_2015-01_revised-minimal}
\bibliography{short_names,biblioli}
\end{document}